	\documentclass[%
	a4paper,10pt,journal,twoside,twocolumn%
	]{IEEEtran}
	\usepackage[square,comma,compress,numbers]{natbib}
	\usepackage[english]{babel}  % avstavning (?)
		\let\ORGforeignlanguage\foreignlanguage
		\def\foreignlanguage#1{\lowercase{\ORGforeignlanguage{#1}}} 
	\usepackage[latin1]{inputenc}
	\usepackage[dvips%
	]{graphicx}
	\usepackage{amssymb}
	\usepackage{amsmath}
	\usepackage{epsfig}
	\usepackage{color} 
	\usepackage[raggedright,tight,hang]{subfigure}
	\usepackage{booktabs}
	\usepackage{srcltx}
	\usepackage{soul}

	\newcommand{\maybespace}{\ }
	
	\newcommand{\maybehalffigure}{\columnwidth}
\newcommand{\maybehighlight}[1]{#1}
\newcommand{\hh}{\ensuremath{\mathfrak{h}}}
\newcommand{\vr}{\vx}
\newcommand{\vx}{\ensuremath{\myvecx}}
\newcommand{\myvec}[1]{\vec{#1}}
\newcommand{\myvecx}{\myvec{r}}

\newcommand{\HH}{\protect{\ensuremath{\mathrm{H}}}}
\newcommand{\LL}{\protect{\ensuremath{\mathrm{L}}}}

\newlength{\figwidthD}

\begin{document}
\title{%
\bf%
%{SURF reverberation suppression transmit-beam generation and post-processing adjustment in an aberrating medium}}
{SURF imaging beams in an aberrative medium: generation and postprocessing enhancement}}
\author{Sven~Peter~{Näsholm},~\IEEEmembership{Member,~IEEE} and Bjørn~A.~J.~{Angelsen},~\IEEEmembership{Senior~Member,~IEEE}%
%\email{peter.nasholm@ntnu.no}

\thanks{%Manuscript received Month DD, 2011; accepted Month DD, 2012. %}
%\thanks{%
This work was supported by the Medicine and Health program of the Research Council of Norway.}
\thanks{The research was mainly done while both authors were with the Department of Circulation and Imaging, Norwegian University of Science and Technology, Trondheim, Norway. Sven Peter Näsholm is now with the Department of Informatics, University of Oslo, Norway (e-mail: svenpn@ifi.uio.no).}
\thanks{Digital Object Identifier http://10.1109/TUFFC.2012.2494} 
}

%\markboth{}{}
%\markboth{\sc IEEE Transactions on Ultrasonics, Ferroelectrics and Frequency Control, Vol.~X, no.~X, Month 201X}{\sc Näsholm and Angelsen: {SURF} imaging beams in aberrating medium}
\markboth{\sc E-print. IEEE Trans.\ Ultrason., Ferroelectr., Freq.\ Control, Vol.~59, no.~11, November 2012}{\sc Näsholm and Angelsen: {SURF} imaging beams in aberrating medium}

\pubid{0000--0000/\$25.00 \copyright\ 2012 IEEE}
%\pubid{\copyright 2011}
\maketitle
\newcommand{\linedescriptions}{%
\includegraphics{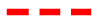}: time-shift adjustment  (homog.),  
\includegraphics{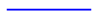}: time-shift (inhomog.), %
\includegraphics{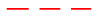}: filter $\hh_{z_a}$ adjustment (homog.), % 
\includegraphics{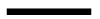}: filter{ }$\hh_{z_a}$ adjustment (inhomog.)%
}
\newcommand{\linedescriptionscoresoverviewfields}{%
\includegraphics{thick_line}: fundamental (inhomog.),%
\raisebox{-.02cm}{\includegraphics{thin_dashed_line_red}}: SURF (inhomog..), %
\includegraphics{thin_blue_line}: fundamental (homog.), %
\raisebox{-.02cm}{\includegraphics{thick_dashed_line_red}}: SURF (homog.).%
}%

\newcommand{\linedescriptionssecfig}{%
	Time-shift adjusted: \includegraphics[width=.6cm]{thick_line} (homogeneous) and \includegraphics[width=.9cm]{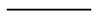} (inhomogeneous. %
	Filter-adjusted: \raisebox{-.03cm}{\includegraphics[width=1cm]{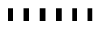}} (homog.) and \raisebox{-.03cm}{\includegraphics[width=.9cm]{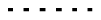}} (inhomog.), %
	No adjustment: \raisebox{-.03cm}{\includegraphics[width=.8cm]{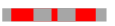}} (homog.) and  \includegraphics[width=.8cm]{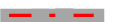} (inhomog.), %
	Fundamental beam: \raisebox{-.02cm}{\includegraphics[width=.9cm]{thick_dashed_line_red}} (homog.) and \raisebox{-.02cm}{\includegraphics[width=.9cm]{thin_dashed_line_red}} (inhomog.). %
	}

{\footnotesize%
\begin{abstract}
This paper presents numerical simulations of dual-frequency second-order ultrasound field (SURF) reverberation suppression transmit-pulse complexes. Such propagation was previously studied in a homogeneous medium. Here instead the propagation path includes a strongly aberrating body-wall modeled by a sequence of delay-screens. %
The applied SURF transmit pulse complexes each consist of a high-frequency imaging 3.5\,MHz pulse combined with a low-frequency 0.5\,MHz sound speed manipulation pulse. %
Furthermore, the feasibility of two signal post-processing methods are investigated using the aberrated transmit SURF beams. These methods are previously shown to adjust the depth of maximum SURF reverberation suppression within a homogeneous medium. %
The request of the study arises because imaging situations where reverberation suppression is useful are also likely to produce pulse wave-front distortion (aberration). Such distortions could potentially produce time-delays that cancel the accumulated propagation time-delay needed for the SURF reverberation suppression technique. %
Results show that both the generation of synthetic SURF reverberation suppression imaging transmit-beams, and the following post-processing adjustments, are attainable even when a body-wall introduces time-delays which are larger than previously reported delays measured on human body-wall specimens.%
\\

\noindent\verb#The peer-reviewed version of this paper is published #\\
\verb#in IEEE Transactions on Ultrasonics, Ferroelectrics #\\
\verb#and Frequency Control, vol. 59, no. 11, pp. 2588-2595,#\\
\verb#November 2012. DOI: 10.1109/TUFFC.2012.2494 #\\
\verb#The final version is available online at #\\
\verb#http://dx.doi.org/10.1109/TUFFC.2012.2494 The current# \\
\verb#e-print is typeset by the authors and differs in e.g.#\\
\verb#pagination and typographic detail.# 

\end{abstract}%
}
%
%
%
%\insertkeywords

\IEEEpeerreviewmaketitle

%-------------------------------------------------------------------
\section{Introduction}
%-------------------------------------------------------------------

%XXXX also cite: 
%-- Yan and Hamilton JASA 2011 statistical model of clutter suppression in tissue harmonic imaging

%\PARstart{T}{his} work concerns dual-frequency transmit-beams that are utilized for reverberation noise suppression in ultrasound image reconstruction by use of methods analyzed in \cite{nasholm2009,nasholm2011}. The developments of the current paper regard the feasibility of such beam generation within an inhomogeneous medium that generates sound-speed variations that aberrate the beam. \maybehighlight{Although the subject is connected, here we don't apply or develop aberration correction algorithms. }%
This work concerns dual-frequency transmit-beams that are utilized for reverberation noise suppression in ultrasound image reconstruction by use of methods analyzed in \cite{nasholm2009,nasholm2011}. The developments of the current paper regard the feasibility of such beam generation within an inhomogeneous medium that generates sound-speed variations that aberrate the beam. \maybehighlight{Although the subject is connected, here we don't apply or develop aberration correction algorithms. }%

{The} quality of medical ultrasound images varies greatly between patients. Spatial inhomogeneities in compressibility and density within the ultrasound propagation path cause additive reverberation (multiple scattering) noise, as well as resolution degradation due to wave-front distortion (aberration). This is especially prominent when imaging through {e.g.} the abdominal wall or the chest wall of obese patients\maybespace\cite{hinkelman:pulse_dist_meas-98, keogh:real_or_artifact-01, mahmutyazicioglu:transabdominal-05, scanlan:artifacts-91, shmulewitz:quality_factors}. %
A recent simulation study supports the hypothesis that the image degradation due to reverberation noise is more prevalent than degradation due to aberration noise \cite{Pinton2011,Pinton2011erratum}. %

In dual-frequency second-order ultrasound field (SURF) reverberation-suppression imaging, a synthetic transmit-beam is generated from the difference between two high-frequency (HF) imaging pulses transmitted in the same direction. %
This beam is in the following simply denoted the SURF beam. %
Both transmissions comprise a dual-frequency band pulse-complex with the HF part used for image reconstruction, added to a low-frequency (LF) part used for material compressibility manipulation. %
The high frequency is typically of the order 10 times the low frequency. %
The compressibility is pressure-dependent due to material nonlinearity. This makes the sound-speed pressure-dependent, being higher for the compressed medium than for the expanded medium \cite{angelsen:ultrasoundbook_II}. %
The LF polarity is switched for the second transmission, therefore making the LF pressure experienced by the second HF pulse opposite what was experienced by the first HF pulse. %
%
%Because they are transmit in conjunction with opposite LF pulse polarity, 
{The pressure-dependent sound-speed thus causes the two HF pulses to propagate at different speeds. %
With increasing distance traveled, the difference in propagation-time required for the two HF pulses hence increases.} %
%
%
% needed in second column of first page if using \pubid
% -------
\pubidadjcol
% -------
%
%
%
%
%
{Consequently, the difference obtained when subtracting the HF pulses is zero at zero depth and grows with covered distance to attain a maximum at 180$^\circ$ phase-shift.} This is the effect that is exploited for reverberation suppression as follows. %
The SURF beam is the synthetic transmit-beam obtained when subtracting the propagated HF pulses at each spatial position. %
{The amplitude of scattered (or reflected) pulses are severely reduced, hence making the nonlinear sound-speed manipulation negligible. Therefore scattered HF parts accumulate less relative time-shift than forward-propagating ones.} %
Especially when the first scattering takes place at shallow depths, multiple scattered contributions within the received HF pulses thus give negligible contribution to the receive SURF difference HF field. 
%
%
%Following the definitions of\maybespace\cite{nasholm2009}, such reverberations are within the Class I and Class II categories and the SURF beam illustrates the reverberation suppression ability by being reduced near the transducer. The tissue harmonic imaging transmit-beam also possesses this property\maybespace\cite{kaupang2011,thomas:thi_why_does_it_work}. %
%
%
%
The reverberation suppression ability of SURF imaging is illustrated by the fact that the beam is reduced near the transducer, as was studied in \cite{nasholm2009} %
where nonlinear numerical propagation simulations in a homogeneous medium were reported. %
In addition, the combination of tissue-harmonic pulse-inversion with SURF reverberation suppression was introduced. %
%In \cite{nasholm2011}, a post-processing enhancement method for SURF reverberation-suppression imaging was presented. %

In case the dominating reverberation noise is due to a strong first scattering (reflection) taking place around a depth $z_a$, the SURF synthetic transmit field may without modification or re-transmission of the sent-out pulses, be adjusted to increase the transmit-beam suppression around $z_a$ by use of post-processing methods as described and analyzed in %
%\cite{nasholm:SURF_postpr} %
\cite{nasholm2011} for a homogeneous medium. %
Both multiple scattering and multiple reflection image artifacts are in the following considered as reverberation artifacts. %
The recent publications \cite{Hansen2011,Masoy2010,hansen2010} concern SURF dual-frequency acoustics within a wider context. %Hansen2011B,
Dual frequency band pulse complexes have also been used for contrast agent detection \cite{deng:dual_frequency, angelsen_hansen:ieee_proc07, bouakaz:radial_modulation, masoy:SURF_in_vivo, hansen2009, emmer2009}.

The purpose of the computer simulation study presented in the current paper is to evaluate the feasibility of SURF transmit reverberation suppression beam generation when a strongly aberrative body-wall is present within the wave-propagation path. Furthermore, it intends to compare these SURF beams to standard fundamental imaging transmit-beams, and in addition to test the post-processing SURF transmit field enhancement techniques described in \cite{nasholm2011} in case of aberration. %

The study is important because the body-wall of a patient where standard fundamental ultrasound imaging is aggravated due to reverberation noise, is also likely to produce propagation time-shifts due to aberration. If these time-shifts heavily distort the accumulated time-shifts needed for adequate SURF synthetic beam generation, the reverberation suppression gain of the SURF imaging method is no longer obtained. %

This paper is organized as follows: First the dual-frequency SURF imaging and fundamental imaging excitation pulses used in the numerical simulations are presented. Then a description of the body-wall model and field simulation method is given. %
The results section show{s} comparisons between transmit-beams from the described excitation pulses both for a homogeneous medium and after propagation through the body-wall. Beam profiles in the focal plane are derived from the transmit-beams, to quantify the effect the body-wall has on the transmit field.
Finally, post-processing adjusted SURF transmit-beams are constructed for some chosen example depths of decreased transmit-beam amplitude. %

%The same raw simulated transmit field data is used as for the 3.5\;MHz setup in \cite{nasholm:SURF_beams}. The additional processing is applied in order to construct $N$ SURF transmit-beams tailored to suppress multiple scattering noise from a set of $N$ near-field depths.

%-------------------------------------------------------------------
\section{Theory and Methods}\label{section:theory_methodsD}
%-------------------------------------------------------------------
%-------------------------------------------------------------------
\subsection{Excitation pulses}
The two excitation pulses utilized to generate the SURF synthetic transmit-beam and the transducer apertures are equal as for a setup applied in \cite{nasholm2009,nasholm2011}. The HF imaging part of each pulse complex has a center frequency at $f_\HH=3.5$\,MHz (50\% fractional bandwidth at $-6$\:dB) while the LF part, whose polarity is inverted for the second pulse, has a center frequency at $f_\LL=0.5$\,MHz (25\% bandwidth at $-6$\:dB). %
The HF and LF surface pressure amplitudes are adjusted to keep the Mechanical Index (MI) below 1.9 and are therefore set to $3.5$ and $0.85$\:MPa respectively. %
{The peak negative pressure value applied for MI calculation is read before filtering the SURF pulse complex, hence taking the amplitude contribution of the LF part into account. Therefore the highest MI is observed for the SURF complex of negative LF manipulation polarity.} %
The axisymmetric aperture is focused at 82\:mm with an outer radius for the HF part of $7.1$\,mm and $10$\,mm for the LF part. The HF pulse transmission is delayed {by $\tau_0$}$\: = 0.2$\,$\mu$s compared to the LF, measured at their centers. Fig.~2~in Ref.~\citenum{nasholm2009} illustrates the influence of $\tau_0$ on the transmit SURF pulse complex. %

The excitation pulse utilized to generate the comparison fundamental imaging field, is equal to the HF part of the SURF complexes described above. 

%\subsection{Summasjon av två HF-pulser, som ær olika aberrerte}
%-------------------------------------------------------------------
%assuming LF not aberrated
%vad borde hænda med HF-profilerna
%
%- LF-nivån som går ned på sidorna borde ta ned sidelobene som kommer av aberrationer

%-------------------------------------------------------------------
\subsection{Body-wall model and pulse propagation simulation}
%-------------------------------------------------------------------
The applied forward wave-propagation simulation method solves the nonlinear Khokhlov-Zabolotskaya-Kuznetsov (KZK) wave equation taking attenuation and interaction between the HF and LF parts of the SURF pulse complexes into account\maybespace\cite{frijlink2008, varslot:propagation}. %
The transducer apertures modeled are rotational symmetric. %Therefore the simulations could in case of a homogeneous medium efficiently be performed using an axisymmetric two-dimensional propagation method. %
{Because} a non-symmetric aberrating body-wall is taken into account within this work, all simulations are performed in three spatial dimensions, hence demanding longer computation times and greater storage capacity {than for 2-D rotational symmetric simulations}. %

\begin{figure}[!tb]
\renewcommand{\figurename}{Fig.} % Make figure captions say Fig. instead of Figure.
  \includegraphics[width=\columnwidth]{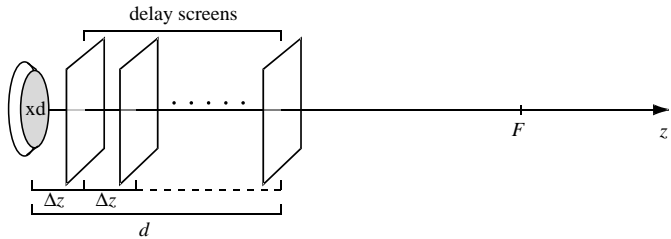}
\caption{Body-wall model and simulation setup outline. The total wall thickness is $d=39$\:mm, the transmit focus is at $F$, and $\Delta z=3$\:mm is the distance separating each delay screen.}
\label{fig:body-wall}
\end{figure}

The body-wall model utilized was generated using a set of 13 two-dimensional filtered time-delay white-noise screens, adjusted to emulate a strongly aberrating abdominal wall. The screens are equally spaced by $\Delta z=3$\,mm thus giving a total wall thickness of $d=39$\:mm as shown in Fig.~\ref{fig:body-wall}. %
This body-wall is similarly modeled as in \cite{varslot:aberr_nonlin_wave_prop}, however the aberrative effect is here chosen to be more severe in order to test the feasibility of the SURF beam generation methods under harsher imaging conditions. The body-wall model construction method is thoroughly described in \cite{masoy:correction-aberration} and the characteristics of the modeled body-wall, as found from simulations of back-propagation of the signal from an excited point source in focus, are displayed in Table \ref{tab:body_wall} %
\begin{table}[!bt]
\renewcommand{\tablename}{\sc TABLE} % Make figure captions say Fig. instead of Figure.
	\caption{Characteristics of the simulated body-wall\label{tab:body_wall}}
	\centering
	\begin{tabular}{l l}
		%\toprule
		%& unit & value\\
		%\midrule
		\toprule 
			Amplitude correlation-length & 1.4 mm\\
			Amplitude RMS fluctuation & 4.5 dB\\
			Time-delay correlation-length& 1.2 mm\\
			Time-delay RMS fluctuation & 450 ns\\
		\bottomrule
	\end{tabular}
\end{table}
At zero-pressure, the material property average values are: speed of sound 1550 m/s, density 1.06 mg/mm$^3$, and the nonlinearity parameter $\beta_n=3.9$. The attenuation is described by a power-law making it proportional to $f^{1.1}$ with the loss $\alpha_0=0.52$\,dB/cm at 1\,MHz. %
{Details on the simulation and attenuation implementations are given in Ref.~}\citenum{abersim_manual}. %

Both the corresponding non-aberrated SURF beam and the non-aberrated fundamental beam are simulated and used as reference beams. %
All beams are generated by detection of the temporal maximum of the pressure field at each {spatial} coordinate. %

%-------------------------------------------------------------------

\subsection{SURF signal post-processing}
%-------------------------------------------------------------------
Two different signal post-processing methods, further described and analyzed for a homogeneous medium in\maybespace\cite{nasholm2011}, are applied on the SURF field data filtered around the frequency $f_\HH = 3.5$\,MHz. %
They both aim to further suppress the SURF transmit-beam at some depth $z_a$, so that the ratio between the beam energy within the imaging depth region and the energy at the chosen suppression depth is increased. This way image reverberation-artifacts due to first scattering from some strongly reflecting feature at the depth $z_a$ may be suppressed. %

In short, given that the two transmit HF waves propagated in conjunction with LF manipulation pulses of opposite polarities are $s_+(\vr,t)$ and $s_-(\vr,t)$, the second HF signal is processed to form the field $\hat s_-(\vr,t)$ which is used to form the difference field
\begin{align}
	s_\Delta(\vr,t) =s_+(\vr,t)-\hat s_-(\vr,t).
\end{align}
The adjusted SURF synthetic transmit-beam is then generated from $s_\Delta(\vr,t)$. %

The first post-processing method involves a pure time-shift of the $s_-(\vr,t)$ field by $\tau_a$ so that 
\begin{align}
	\hat s_-(\vr,t)=s_-(\vr,t-\tau_a).
\end{align}

The second post-processing method involves some more general filter $\hh_{z_a}$ which makes the on-axis pulse of $s_+$ and $s_-$ equal at the depth $z_a$:
\begin{align}
	\hat s_-(\vr,t)=\hh_{z_a}\big\{s_-(\vr,t)\big\}.
\end{align}
Here this filter $\hh_{z_a}$ is determined by comparison of the two simulated fields as averaged over a laterally $0.5$\,mm wide region $S$ on-axis at the depth $z_a$ and finding the filter that makes these two average signals equal. %
For the demonstrated inhomogeneous medium case, the adjustment filter $\hh_{z_a}$ utilized is not generated from the actual aberrated $s_+(\vr,t)$ and $s_-(\vr,t)$ fields, but instead from the $s_+(\vr,t)$ and $s_-(\vr,t)$ emerging from the homogeneous medium propagation. This approach is chosen to in some extent better emulate how an \emph{à priori} determination of $\hh_{z_a}$ %
{could be performed by experimental field characterization}. %
The averaging over the region $S$, which was not performed in \cite{nasholm2011}, is introduced to take a finite hydrophone width into account. % 
 The determination of $\hh_{z_a}$ in a real medical imaging setup may be more cumbersome, as further discussed in \cite{nasholm2011}. %
{We remark that when receive channel-data is available, any of the proposed suppression methods may be applied, and images can be reconstructed independently for multiple choices of $z_a$. Great flexibility in reverberation suppression customization is hence enabled. }

%-------------------------------------------------------------------
\subsection{Compared transmit-beams and fields}
%-------------------------------------------------------------------
In the following is a listing of the simulated datasets utilized to generate the transmit-beams that are compared within this work: %
\subsubsection{Non-adjusted SURF beam\label{subsubsection:non-adjusted}} %
The HF transmit fields $s_+(\vr,t)$ and $s_-(\vr,t)$, filtered around $f_\HH=3.5$\,MHz on the same actual dataset as utilized in \cite{nasholm2009}, are subtracted without further modification to generate $s_\Delta(\vr,t)=s_+(\vr,t) - s_-(\vr,t)$. 
\subsubsection{Time-shift adjusted SURF beams} % 
The time-shift adjustment post-processing method is applied on the same original dataset as used to generate the non-adjusted SURF, for a number of different $z_a$ to generate the adjusted HF field $s_\Delta(\vr,t) = s_+(\vr,t) - s_-(\vr,t-\tau_a)$.
\subsubsection{Filter-adjusted SURF beams} %
The filter-adjustment post-processing method is applied on same original dataset as for the non-adjusted SURF, for a number of different $z_a$ to generate the adjusted HF field $s_\Delta(\vr,t) = s_+(\vr,t) - \hh_{z_a}\big\{s_-(\vr,t-\tau_a)\big\}$.
\subsubsection{Fundamental imaging beam} %
Standard fundamental imaging transmit fields without LF manipulation or SURF processing. The transmit pulse is equal to the HF part of the SURF pulse complex described above. %

%

%-------------------------------------------------------------------
\section{Results and Discussion}
%-------------------------------------------------------------------

%-------------------------------------------------------------------
%\subsection{Non-adjusted SURF fields compared to fundamental fields}
%-------------------------------------------------------------------

\begin{figure}[!tb]
\renewcommand{\figurename}{Fig.} % Make figure captions say Fig. instead of Figure.
\centering
	\includegraphics[width=.92\maybehalffigure]{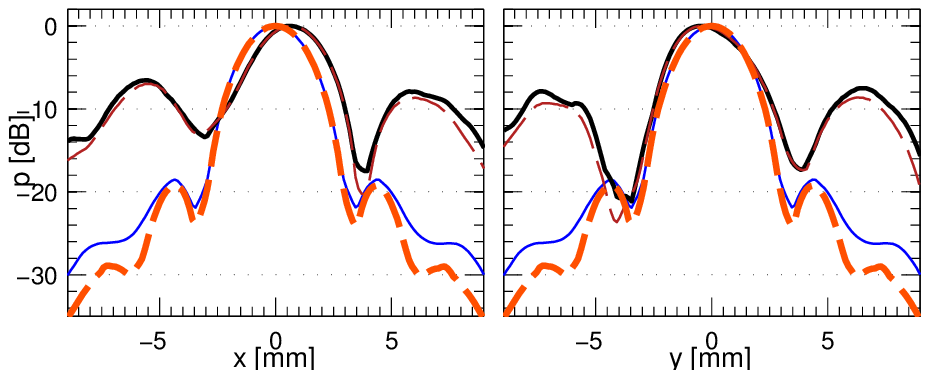}%
	{\smallskip}
	\\%mayberowbreak
	\includegraphics[width=.92\maybehalffigure]{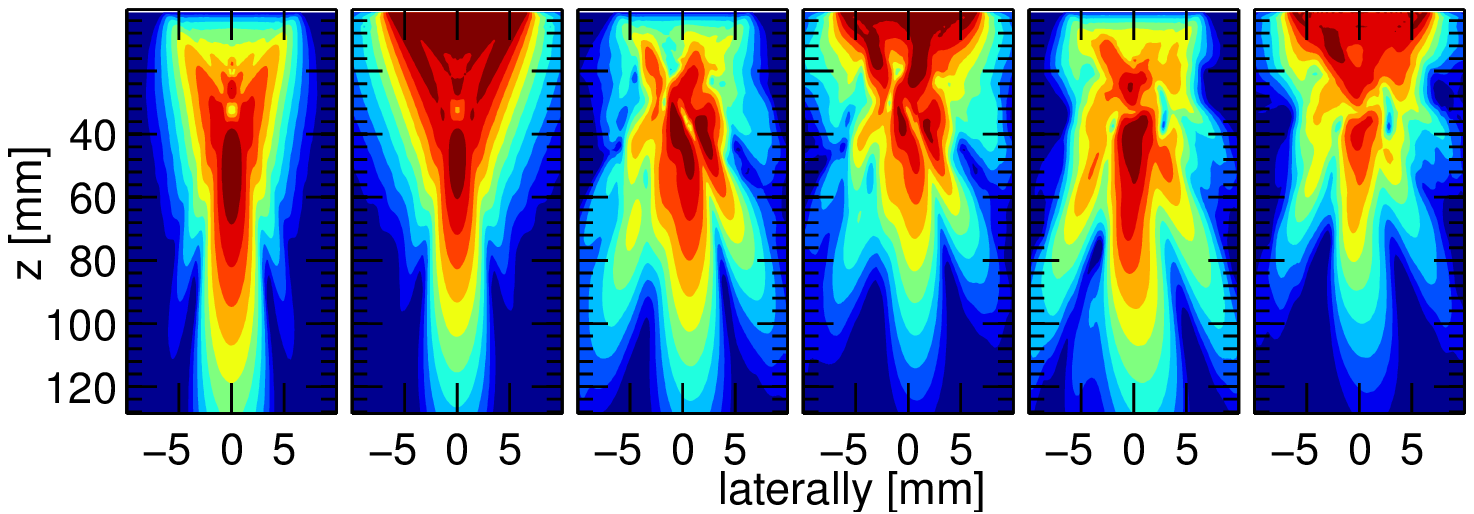}%
	\:\raisebox{.5cm}{\includegraphics[height=2.3cm]{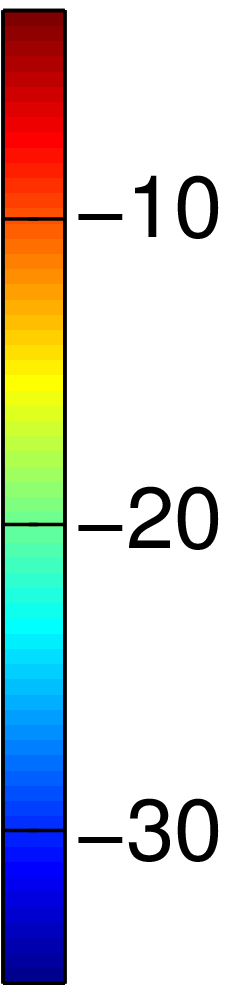}}\quad
	%{\smallskip}
	\\
	\includegraphics[width=.92\maybehalffigure]{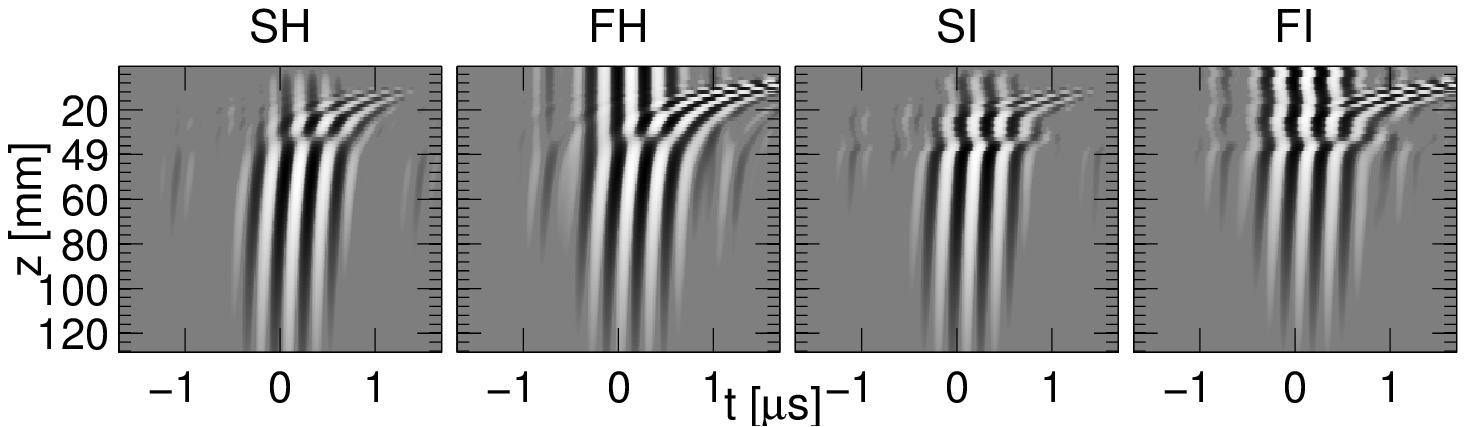}%
	{\smallskip}
	\\%mayberowbreak
	\includegraphics[width=.92\maybehalffigure]{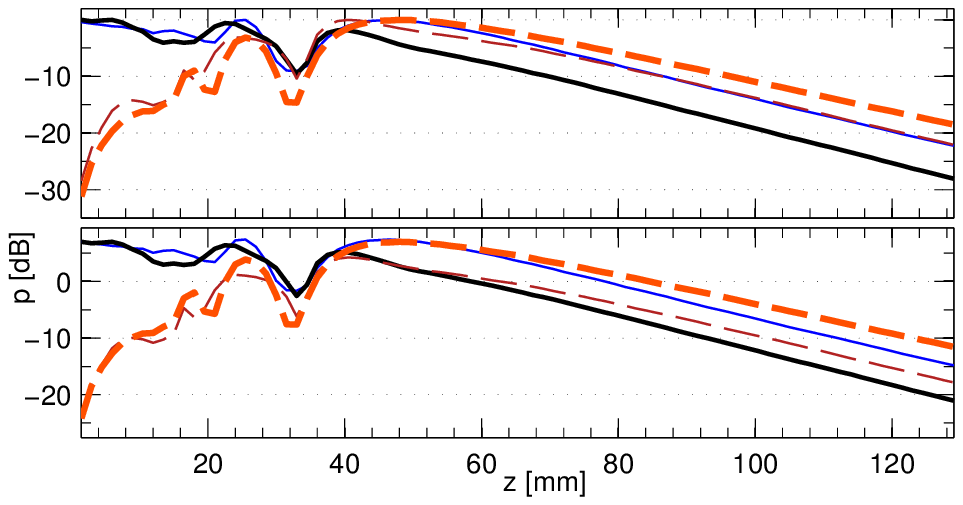}%
	\caption[]{{Body-wall model impact on the transmit beams illustrated through comparison between non-adjusted SURF and fundamental imaging transmit fields, for homogeneous and inhomogeneous medium. %
	Top pane row: beam profiles in focus. On the left along the $x$ axis, on the right along the $y$ axis. %
	Second pane row from top: Transmit-beams, from left to right: SH, FH, SI($zx$), FI($zx$), SI($zy$), and FI($zy$) with the abbreviations explained in Table~}\ref{tab:short_names} %
	($zx$) and ($zy$) refer to two planes perpendicular to the beam axis. %
	Third pane row from top: Beam-axis pulses, from left to right: SH, FH, SI, and FI. %
	Bottom two panes: beam profiles along the depth axis. Individual normalization on top and common normalization below. %
	Line notations: \linedescriptionscoresoverviewfields{ }}
  \label{fig:fields_no_processing}
\end{figure}

\begin{figure*}[!htb]
\renewcommand{\figurename}{Fig.} % Make figure captions say Fig. instead of Figure.
	\ \includegraphics[scale=.6]{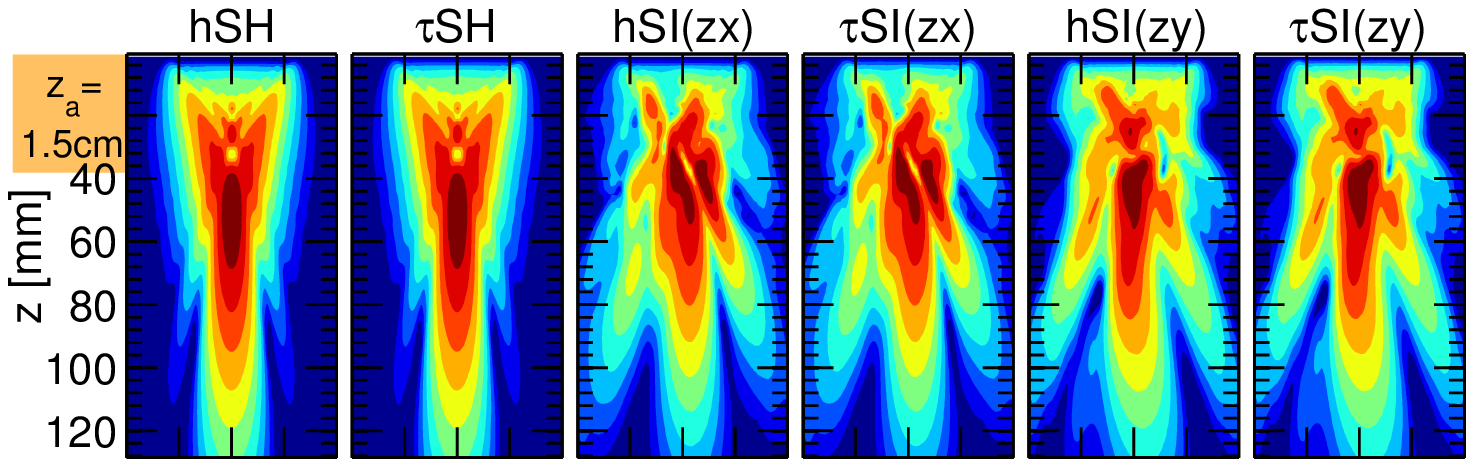}%
	\:\includegraphics[scale=.6]{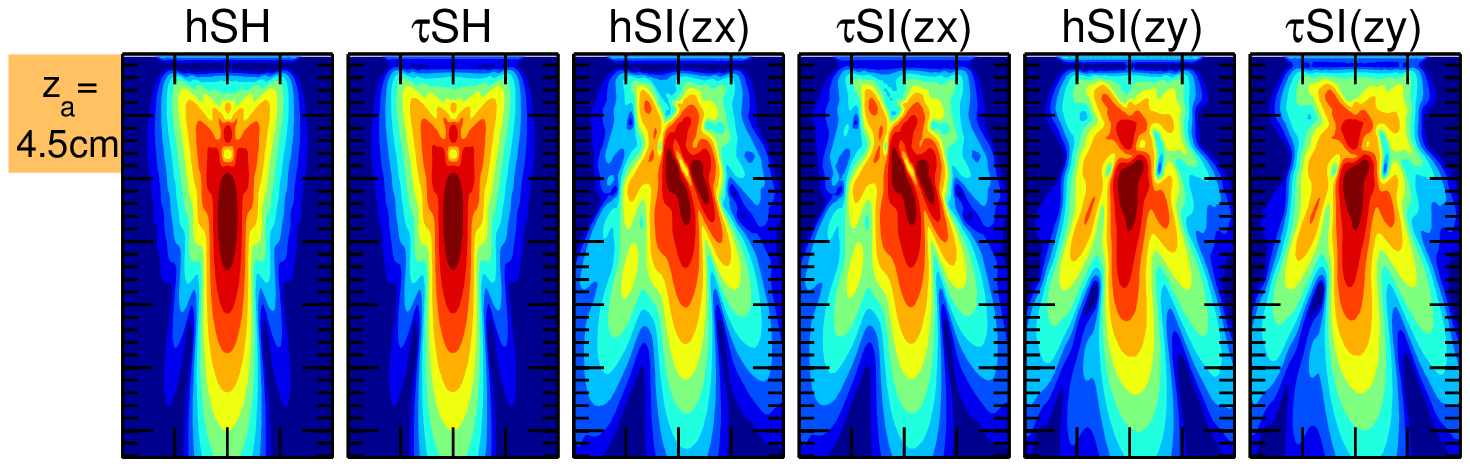}\\%
	\includegraphics[scale=.6]{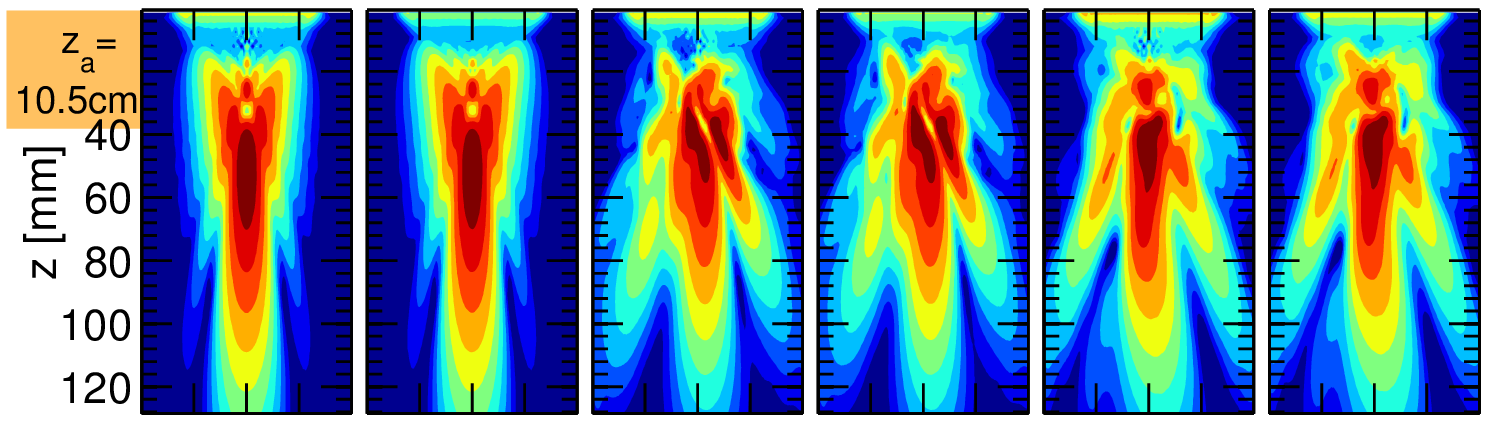}%
	\includegraphics[scale=.6]{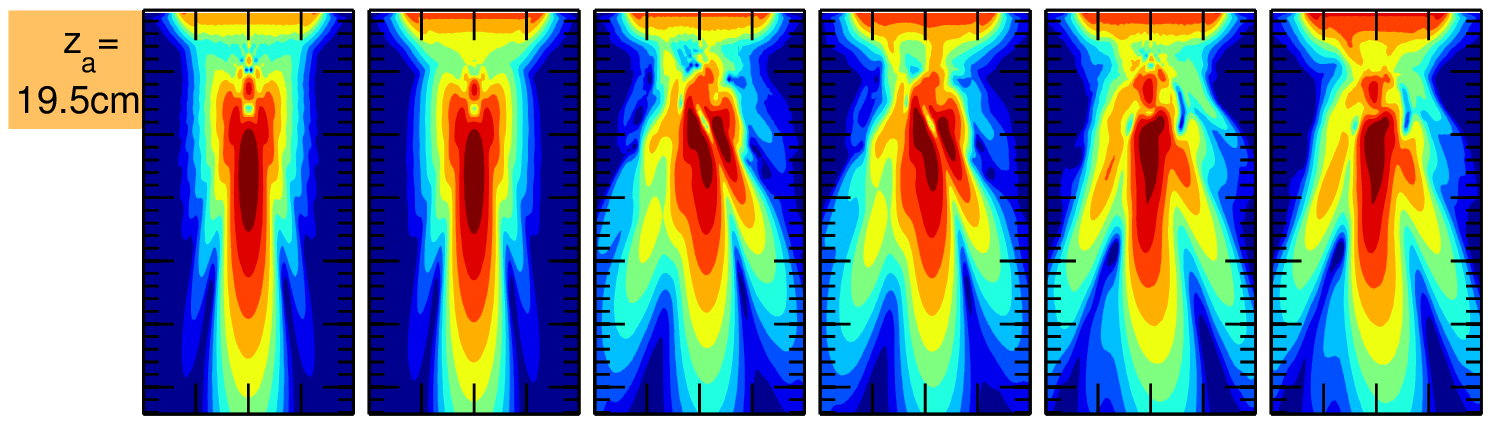}\\%
	\includegraphics[scale=.6]{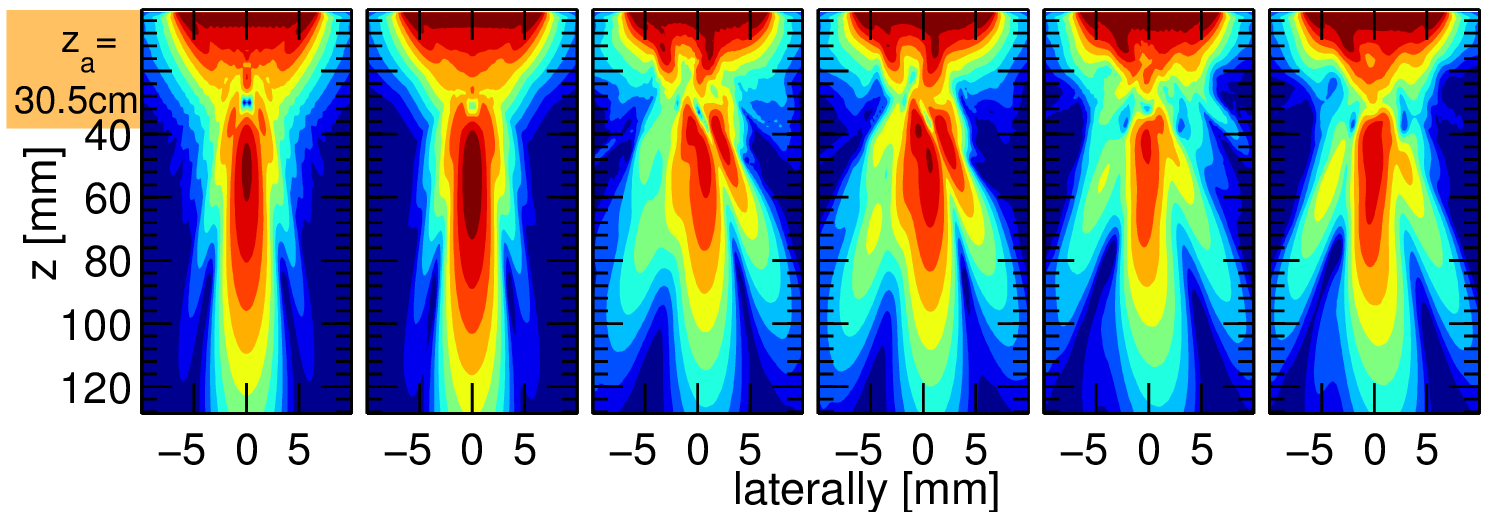}%
	\:\ \includegraphics[scale=.6]{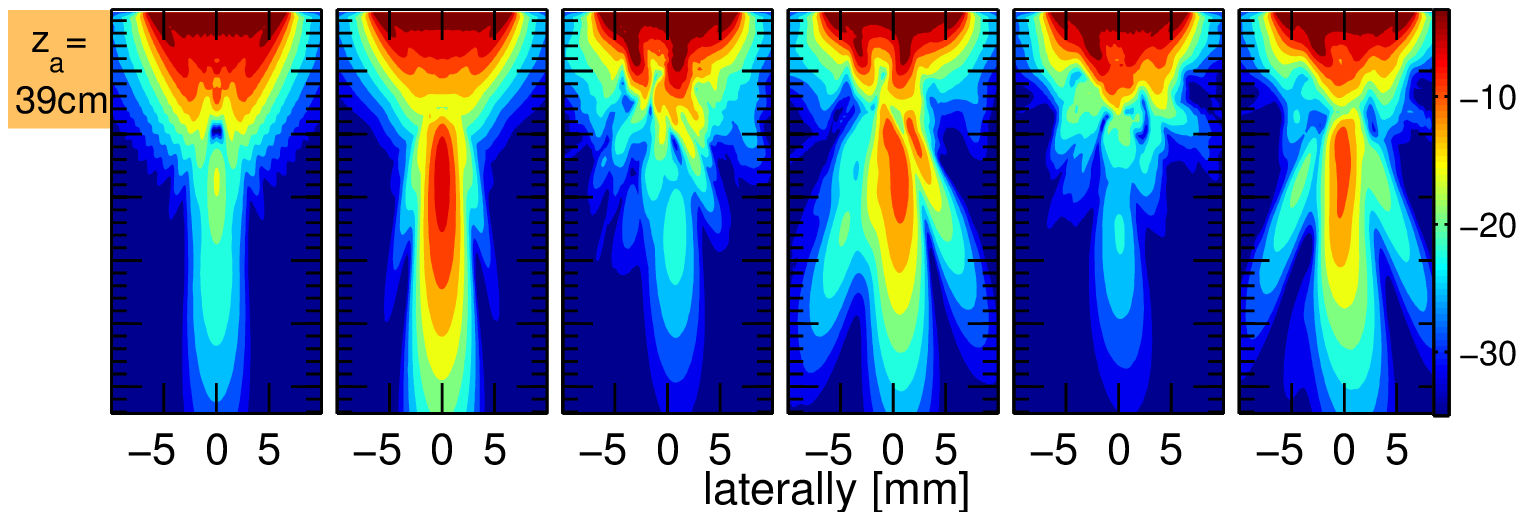}% 
	\caption[]{{Adjusted SURF beam cuts for $z_a=1.5$\:cm (top left pane group), $z_a=4.5$\:cm (top right), $z_a=10.5$\:cm (middle left), $z_a=19.5$\:cm (middle right), $z_a=30.5$\:cm (bottom left), and $z_a=39$\:cm (bottom right). Abbreviation codes shown in Table }\ref{tab:short_names}{  %
	Beams, from left to right within each $z_a$ pane group: $\hh$SH, $\tau$SH, $\hh$S{I}($zx$), $\tau$S{I}($zx$), $\hh$S{I}($zy$), and $\tau$S{I}($zy$). %
	The bottom-right colorbar indicates the 35\:dB dynamic range\maybehighlight{ that is used for all the above beam cuts.}}}
	\label{fig:adjusted_beams}%
\end{figure*}

\begin{figure*}[!htb]
\renewcommand{\figurename}{Fig.} % Make figure captions say Fig. instead of Figure.
	\begin{tabular}{c c c}
	\includegraphics[scale=.6]{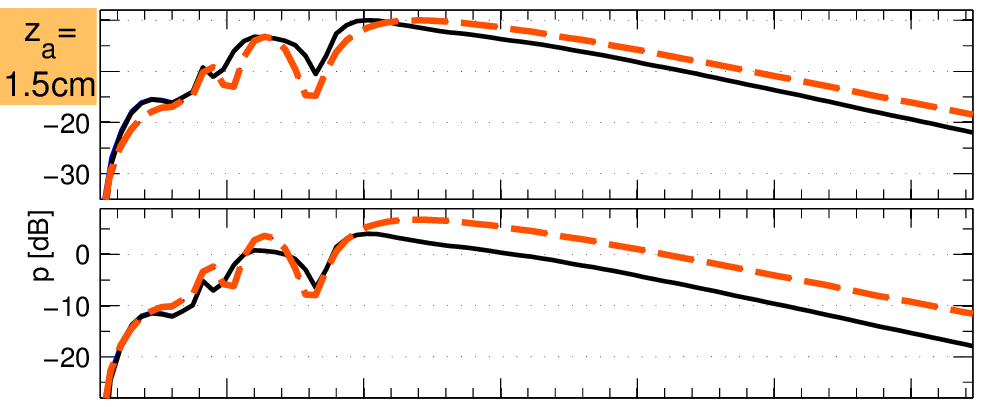}%
	& %
	\includegraphics[scale=.6]{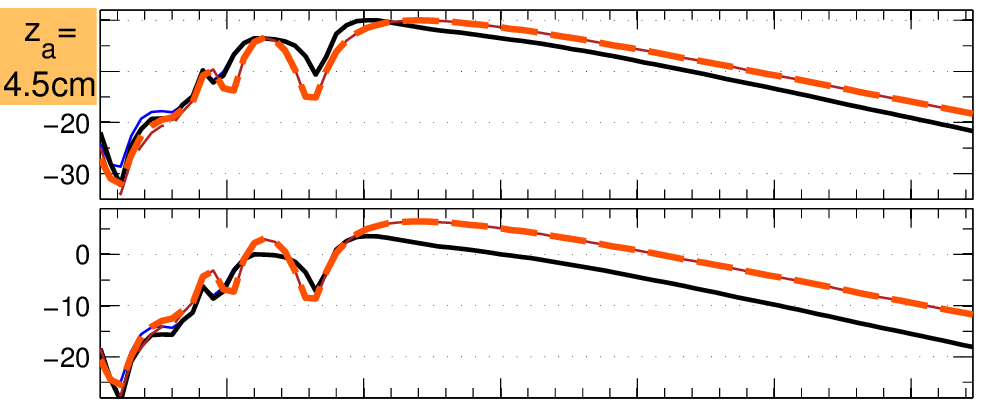}%
	& %
	\includegraphics[scale=.6]{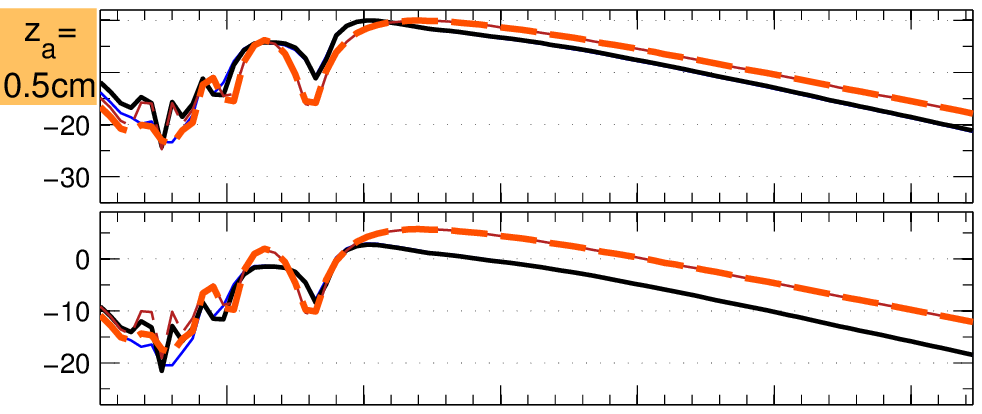}%
	\\%
	\hline%
	\noalign{\smallskip}
	\includegraphics[scale=.6]{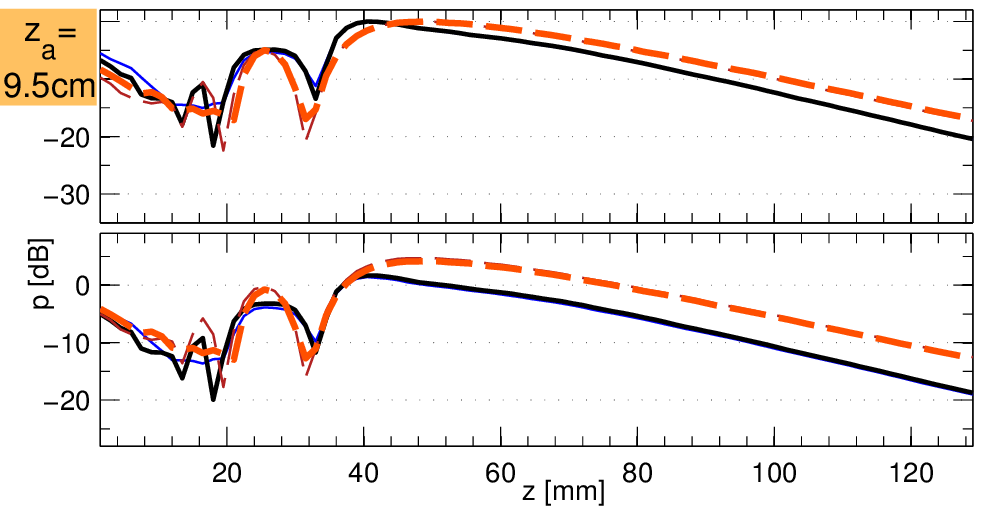}%
	& %
	\includegraphics[scale=.6]{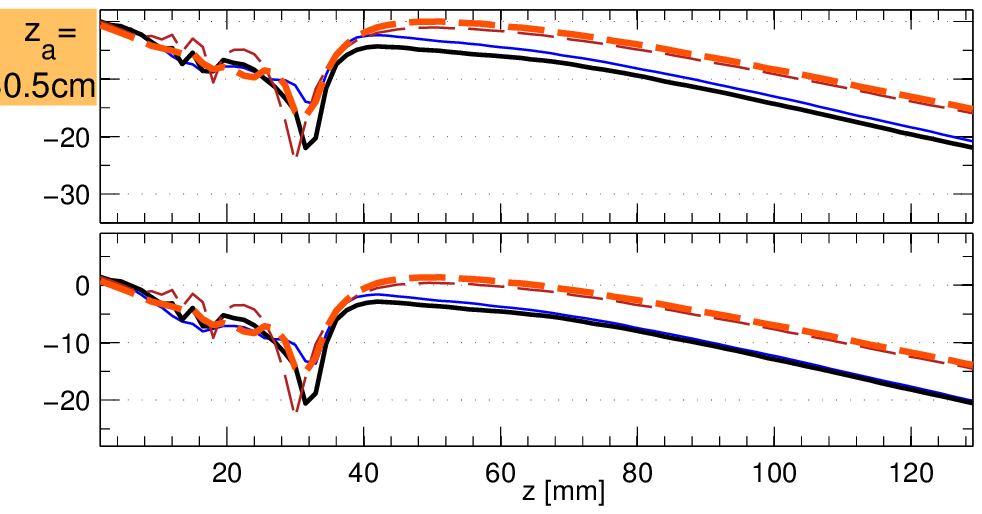}%
	& %
	\includegraphics[scale=.6]{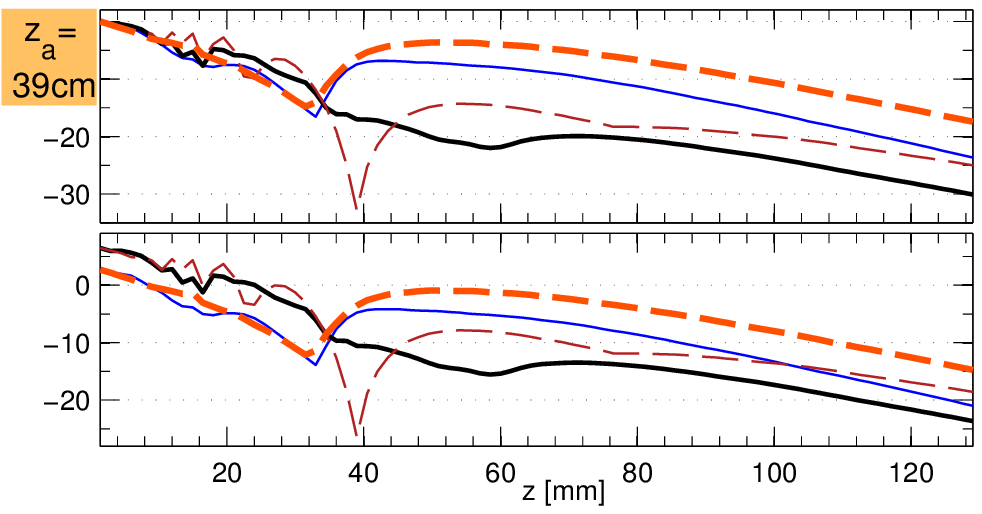}% 
	\end{tabular}
	\caption[]{{On-axis SURF beam profiles as function of depth $z$ for $z_a=1.5$\:cm (top left pane group), $z_a=4.5$\:cm (top right), $z_a=10.5$\:cm (middle left), $z_a=19.5$\:cm (middle right), $z_a=30.5$\:cm (bottom left), and $z_a=39$\:cm (bottom right). 
	Line descriptions: }\linedescriptions
	}
	\label{fig:beam_profiles}%
\end{figure*}

\begin{figure*}[!htb]
\renewcommand{\figurename}{Fig.} % Make figure captions say Fig. instead of Figure.
	\begin{tabular}{c c}
	\includegraphics[scale=.6]{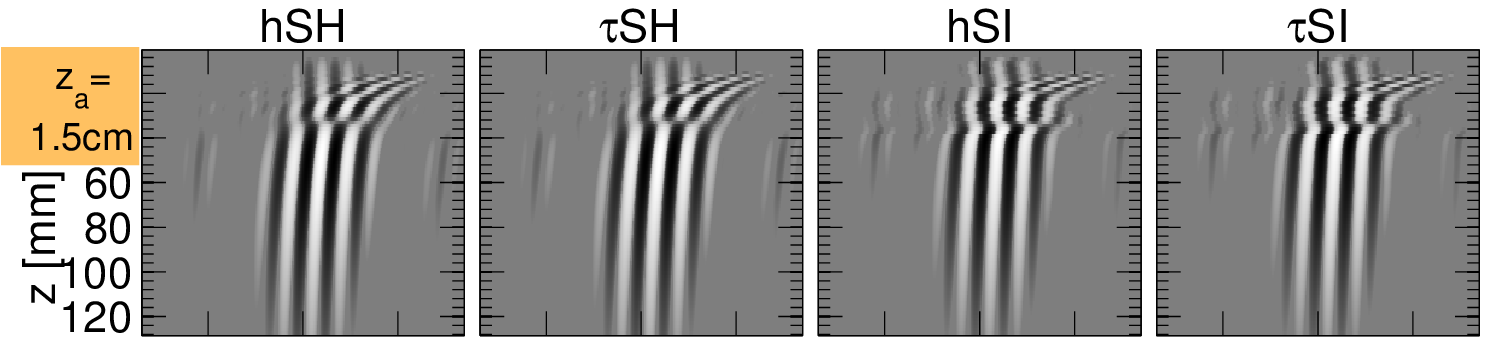}%
	& \includegraphics[scale=.6]{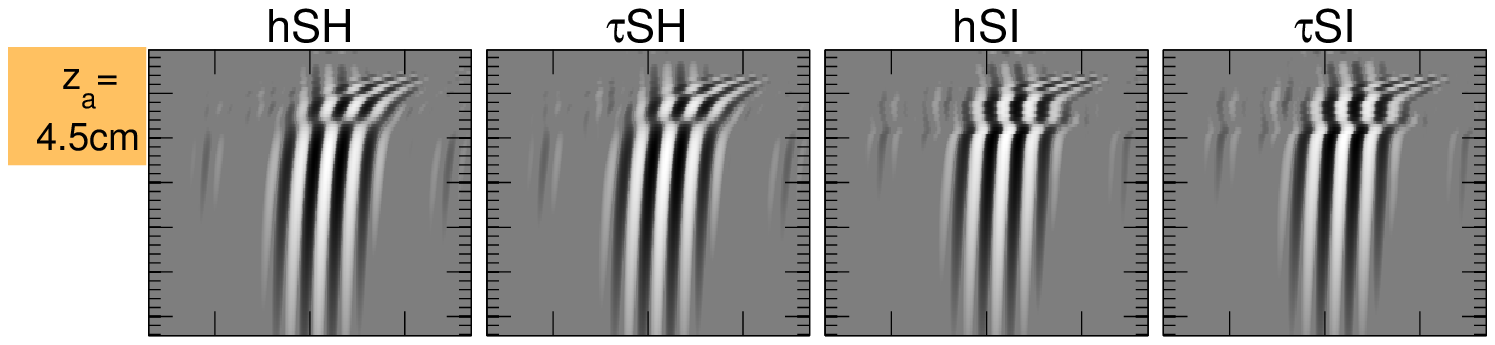}%
	\\%
	\hline%
	\noalign{\smallskip}
	\includegraphics[scale=.6]{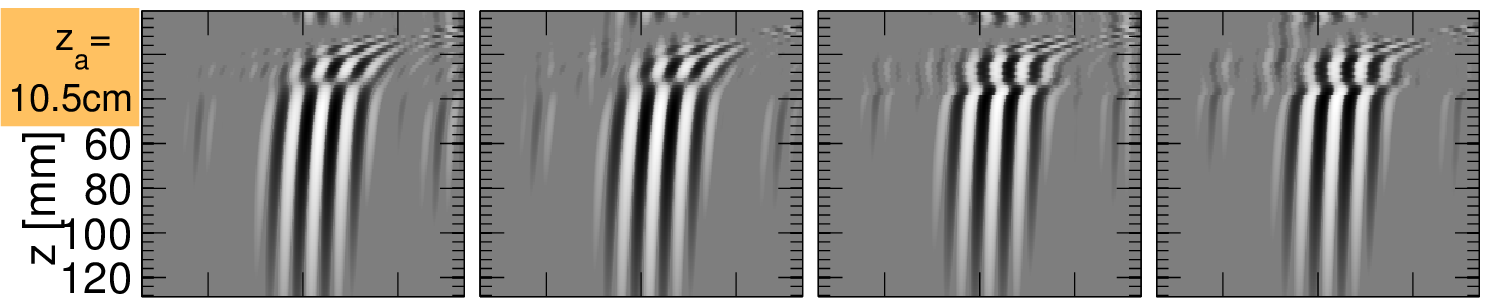}%
	& \includegraphics[scale=.6]{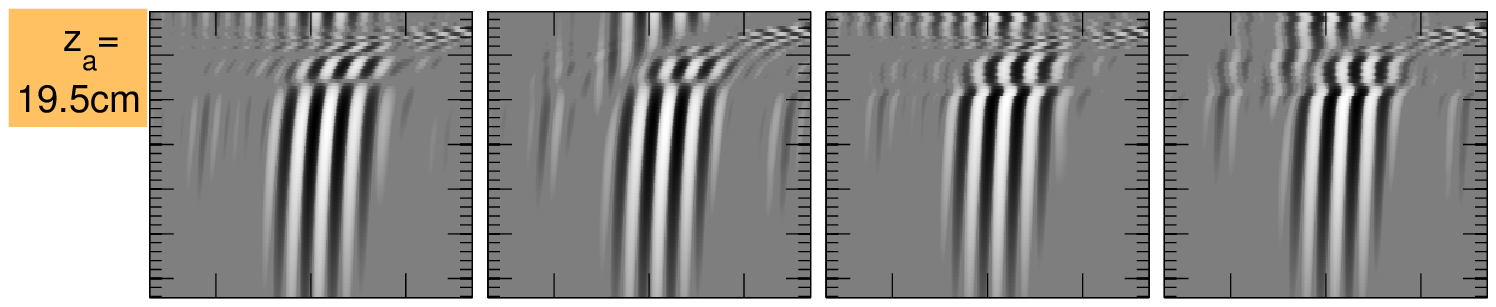}%
	\\%
	\hline%
	\noalign{\smallskip}
	\includegraphics[scale=.6]{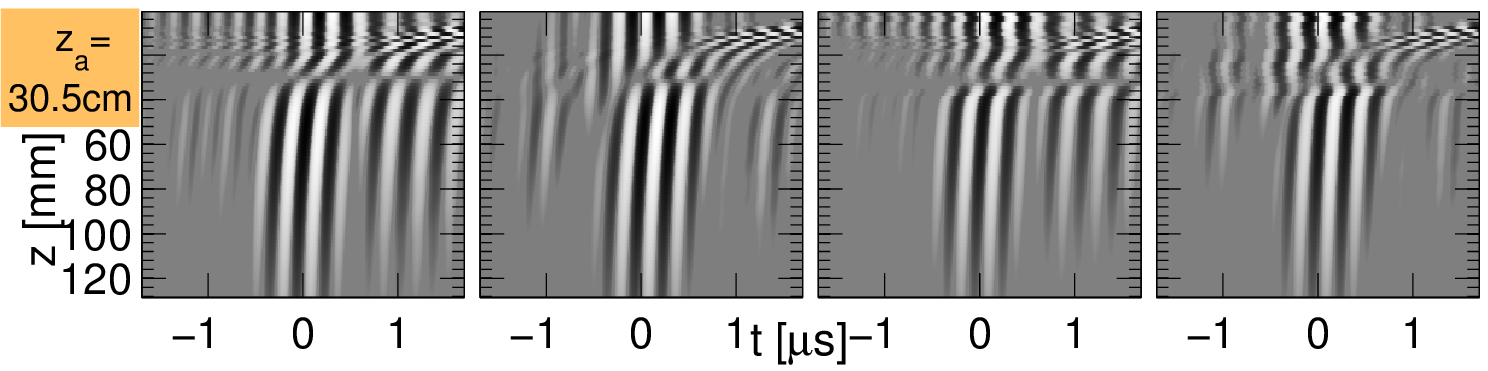}%
	&\ \includegraphics[scale=.6]{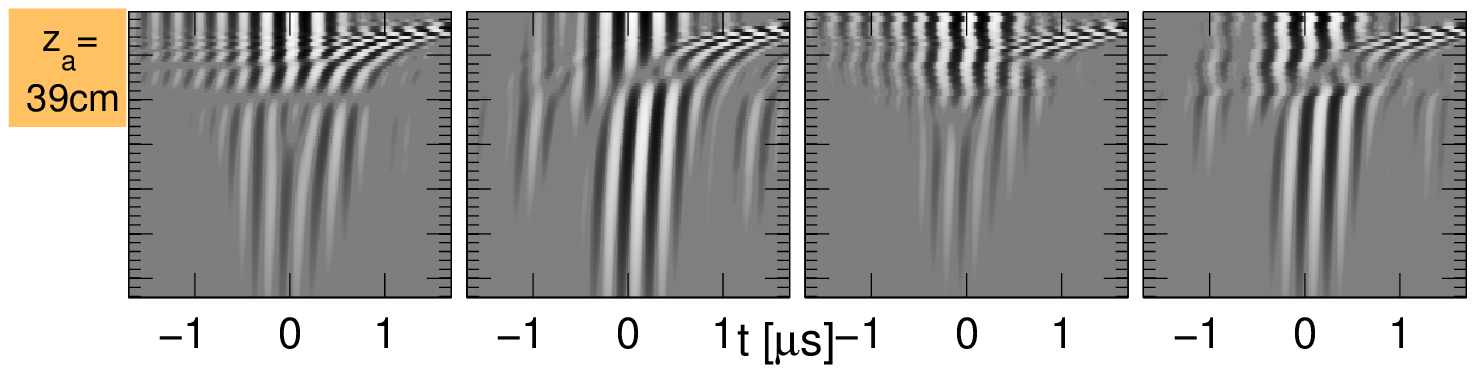}% 
	\end{tabular}
	\caption[]{{On-axis SURF pulses as function of depth $z$ and time $t$ for $z_a=1.5$\:cm (top left pane group), $z_a=4.5$\:cm (top middle), $z_a=10.5$\:cm (top right), $z_a=19.5$\:cm (bottom left), $z_a=30.5$\:cm (bottom middle), and $z_a=39$\:cm (bottom right). Abbreviation codes shown in Table }\ref{tab:short_names}{  %
	Beams, from left to right within each $z_a$ pane group: $\hh$S{H}, $\tau$S{H}, $\hh$S{I}, and $\tau$S{I}. %
	}}
	\label{fig:axis_pulses}%
\end{figure*}

\begin{figure*}[!htb]
\renewcommand{\figurename}{Fig.} % Make figure captions say Fig. instead of Figure.
	\centering
	\begin{tabular}{c c c}
	\includegraphics[width=.31\textwidth]{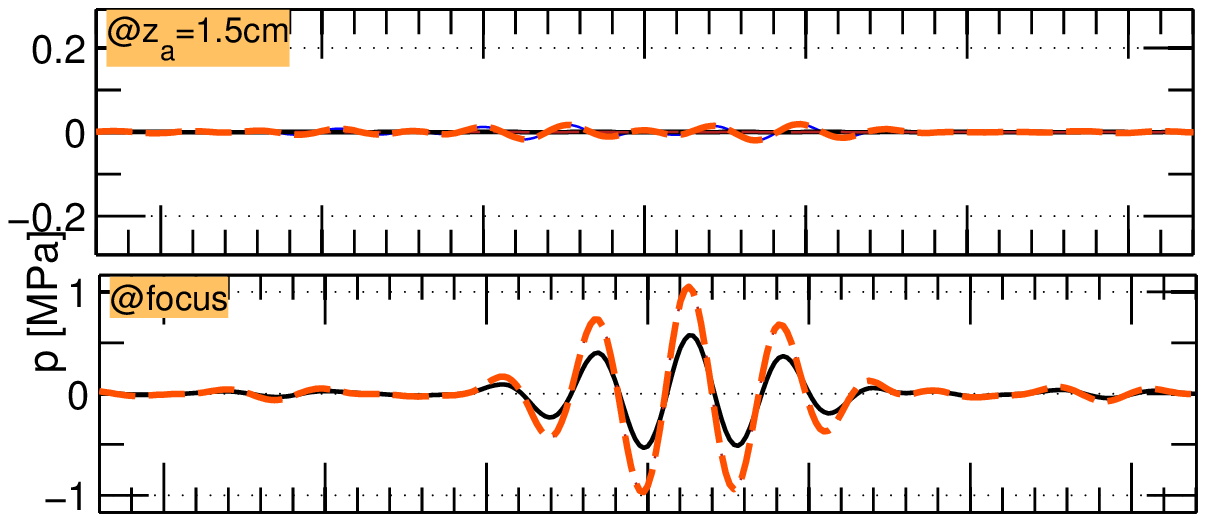}%
	& %
	\includegraphics[width=.31\textwidth]{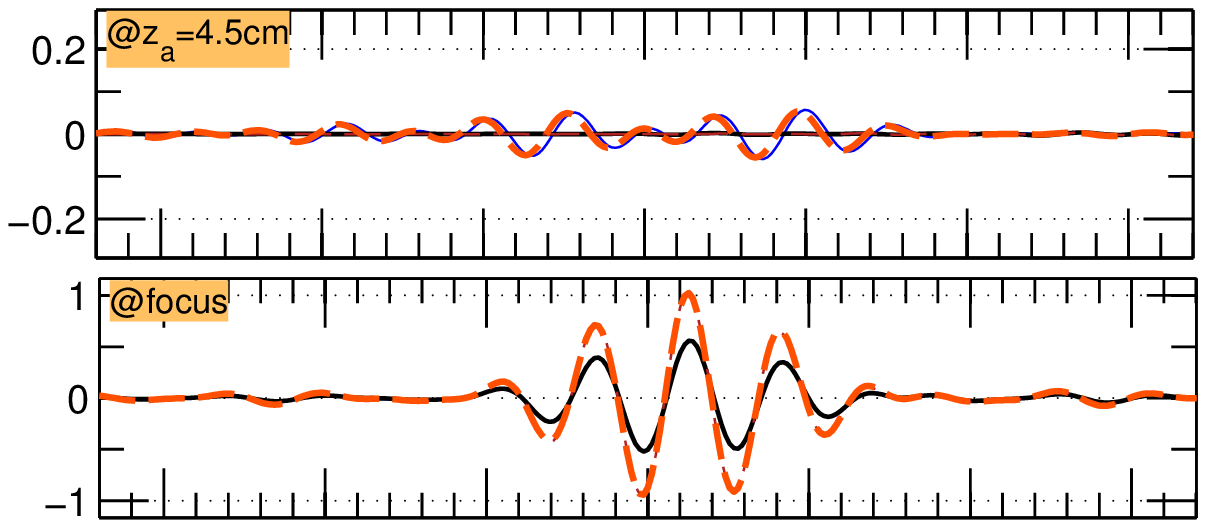}%
	& %
	\includegraphics[width=.31\textwidth]{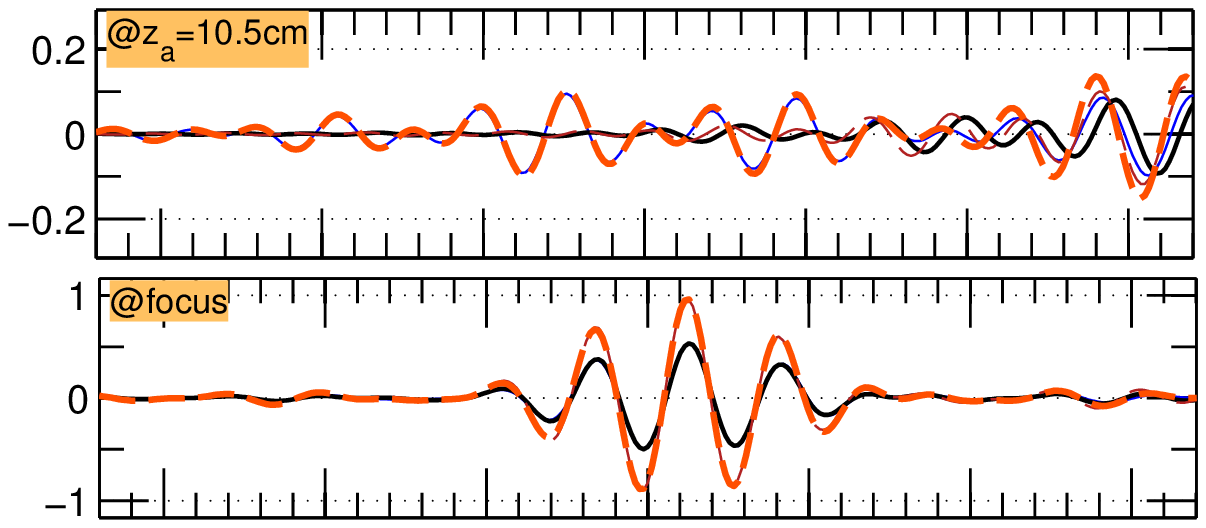}%
	\\%
	\noalign{\smallskip}
	\hline%
	\noalign{\smallskip}
	\noalign{\smallskip}
	\includegraphics[width=.31\textwidth]{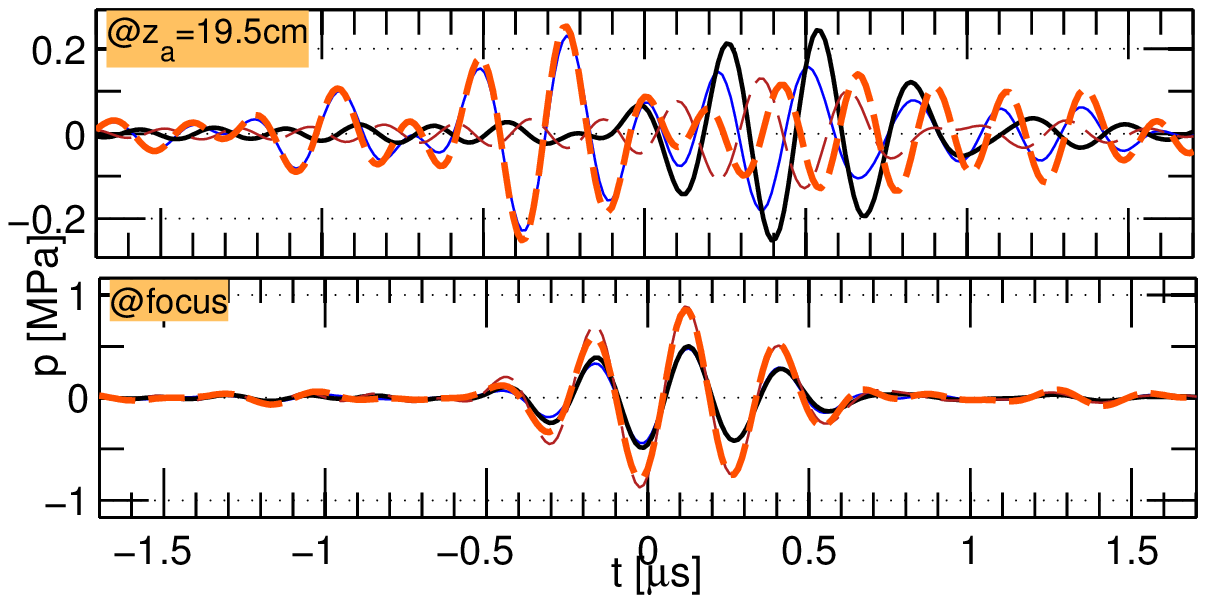}%
	& %
	\includegraphics[width=.31\textwidth]{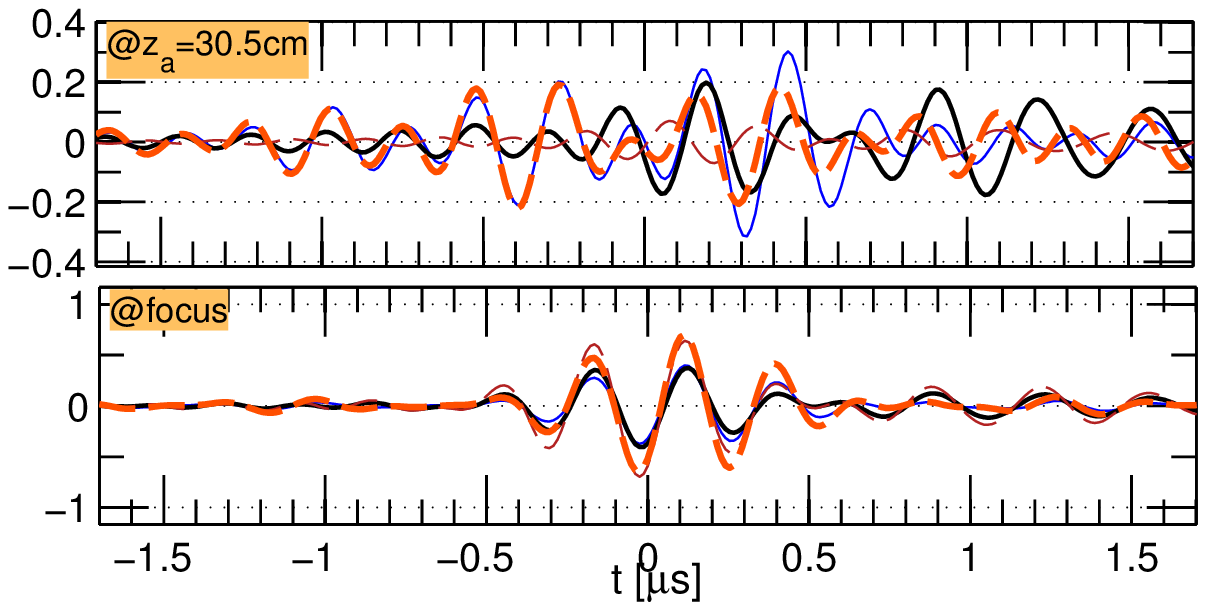}%
	& %
	\includegraphics[width=.31\textwidth]{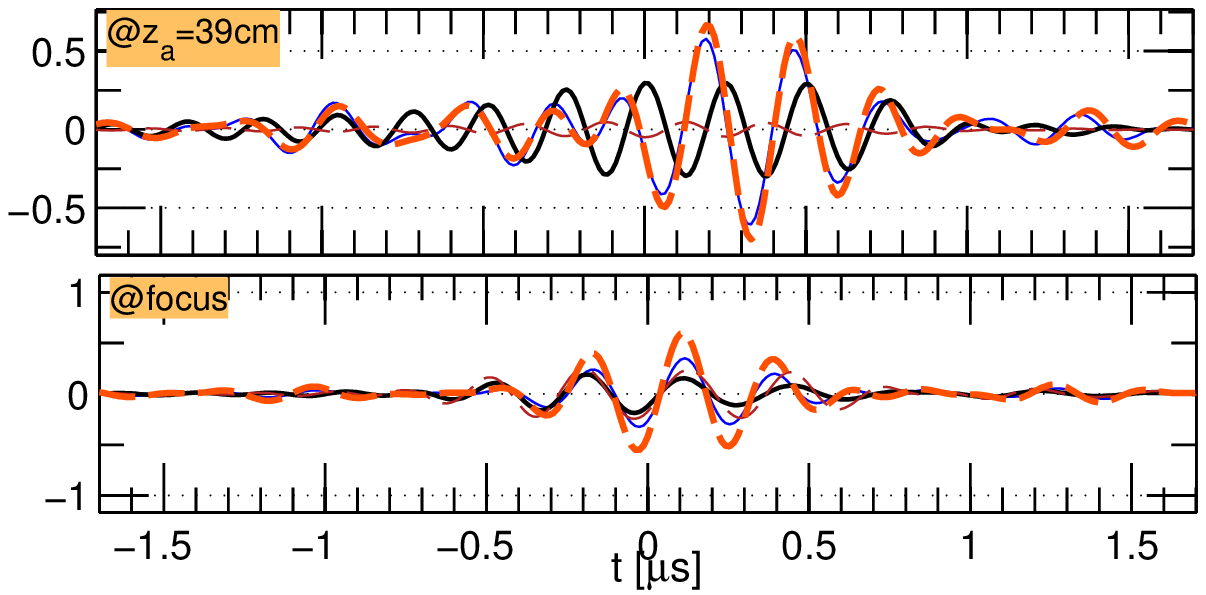}% 
	\end{tabular}
	\caption[]{{On-axis pressure in focus (bottom sub-panes of each pair) and at $z_a$ (top sub-panes of each pair) as function of time $t$ for $z_a=1.5$\:cm (top left pane group), $z_a=4.5$\:cm (top right), $z_a=10.5$\:cm (middle left), $z_a=19.5$\:cm (middle right), $z_a=30.5$\:cm (bottom left), and $z_a=39$\:cm (bottom right). 
	Line notation as in Fig.~}\ref{fig:beam_profiles}.
	}
	\label{fig:time_pulses}%
\end{figure*}

{The propagation effects caused by the modeled body-wall, including focus degradation, are further illustrated in Fig}~\ref{fig:fields_no_processing}, %
which shows beam profiles, transmit-beams, and axial pulses for (non-adjusted) SURF and fundamental beams. % 
\begin{table}[!bt]
\renewcommand{\tablename}{\sc TABLE} % Make figure captions say Fig. instead of Figure.
	\caption{\label{tab:short_names}Abbreviations used in the field descriptions of Fig.~\ref{fig:fields_no_processing} -- \ref{fig:beam_profiles}.}
	\centering
	\begin{tabular}{l c c c c}
		\toprule
			&             &\multicolumn{3}{c}{SURF} \\
			\cmidrule{3-5}
			&             & non- & time-shift & filter \\
			& fundamental & adjusted & adjusted & adjusted\\
		\midrule
			homogeneous & FH & SH & $\tau$SH & $\hh$SH\\
			inhomogeneous & FI & SI & $\tau$SI & $\hh$SI\\
		\bottomrule
	\end{tabular}
\end{table}
%
%
%
%
%-------------------------------------------------------------------
%\subsection{Adjusted SURF fields}
%-------------------------------------------------------------------
Figure \ref{fig:adjusted_beams} shows adjusted SURF beams simulated by homogeneous and inhomogeneous propagation with 6 different choices of suppression depths $z_a$, using both time-shift adjustment or filter-adjustment. The corresponding beam profiles along the $z$ axis are shown in Fig.~\ref{fig:beam_profiles}. %
The time-development of the SURF on-axis pulses as a function of depth are displayed in \ref{fig:axis_pulses} for the same $z_a$ set. These pulses are further inspected in Fig.~\ref{fig:time_pulses}, where their signatures at the $z_a$ depths are compared to the signatures in focus. %
{Side-by-side comparison of the adjusted beams for different $z_a$ is of particular significance, as it visualizes that for the selected $z_a$ depths, the adjustment methods are about as viable for inhomogeneous as for homogeneous media. }%

\begin{figure}[!bth]
\renewcommand{\figurename}{Fig.} % Make figure captions say Fig. instead of Figure.
\centering
	\includegraphics[width=\maybehalffigure]{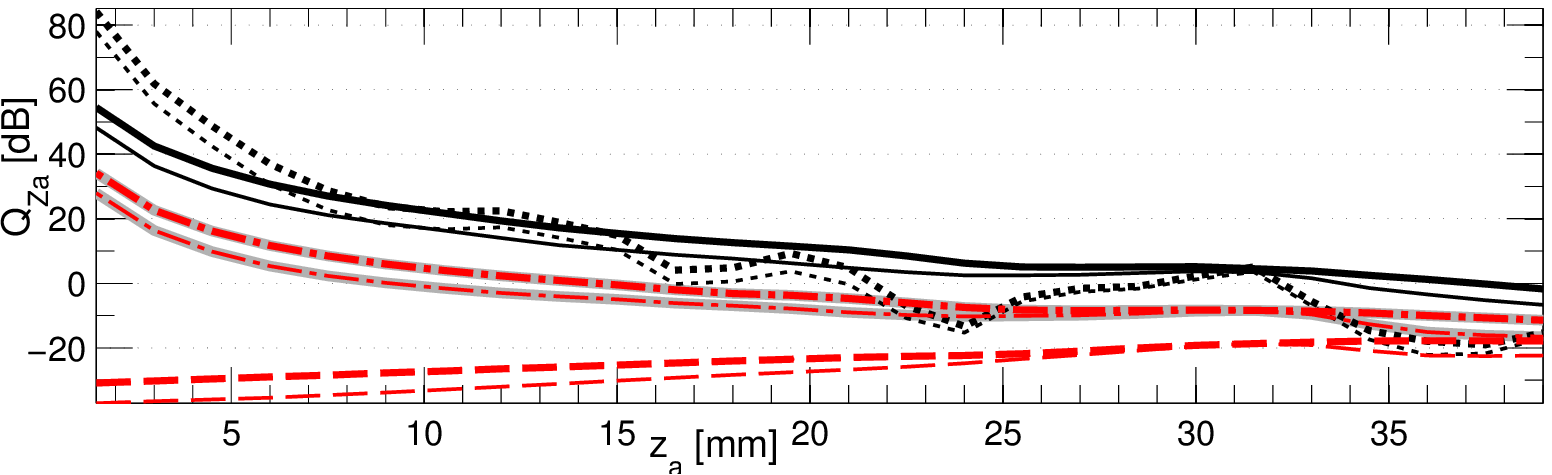}\\ %
	\includegraphics[width=\maybehalffigure]{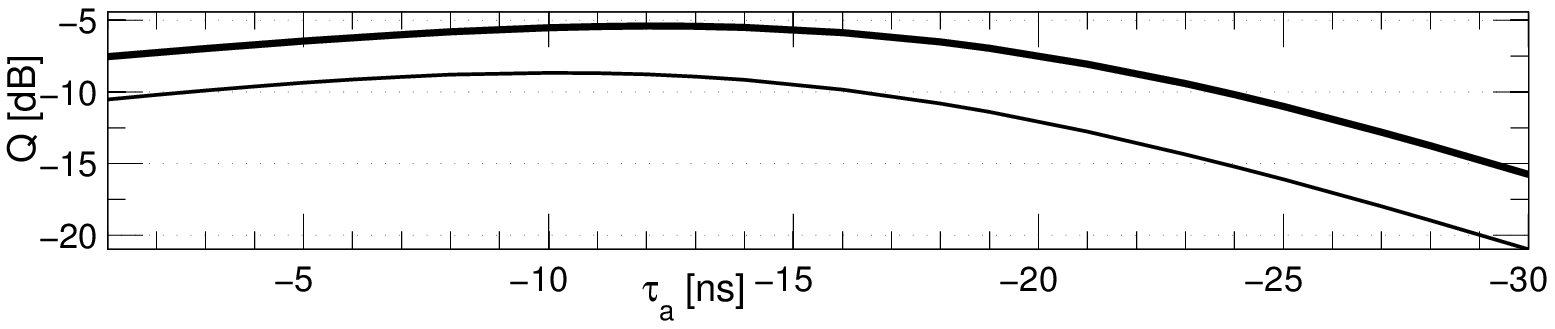}%
	\caption[]{%
	Reverberation suppression beam quality measures with the specific ratios $Q_{z_a}$ {as a function of $z_a$} in the {top} pane and the general $Q$ {as a function of the adjustment time-shift $\tau_a$} in the {bottom} pane. %
	\linedescriptionssecfig{ }
	 }%
	\label{fig:scores}%
\end{figure}

The {adjustment abilities in inhomogeneous media are further quantified by the} specific reverberation suppression beam quality ratios $Q_{z_a}$ corresponding to the shown beams. These are {displayed} for different suppression depths in Fig.~\ref{fig:scores} using time-shift adjustment and filter-adjustment as well as for non-adjusted and fundamental beams. This is done both for homogeneous medium and for propagation through the body-wall. The figure also shows general reverberation suppression quality ratios $Q$ for different time-shift adjustments for the same medium cases. %
%
%
%
%Table \ref{tab:line_notation} shows the line notations used in the figure. %
The specific quality ratio was defined in \cite{nasholm2011} as:%
\begin{align}
	Q_{z_a}\triangleq {\displaystyle \sum_{z=z_n}^{z_f} \sum_{r=0}^{\infty}\sum_{\theta=0}^{2\pi} E(\vx)}  \left/\   {\displaystyle \sum_{r=0}^{\infty}\sum_{\theta=0}^{2\pi} E(\vx)}\big|_{z=z_a}\right.,
	\label{eq:quality_measureD}
\end{align}
where $E(\vx)$ is the beam energy at the spatial coordinate $\vx$ described by the cylindrical coordinates $(z,r,\theta)$. The imaging depth region is within $z\in[z_n,z_f]$ and is equal to the focal region with the given aperture and imaging frequency. %
{The measure $Q_{z_a}$ illustrates how well the transmit beam is adapted to suppression of reverberations where the first scattering takes place at the depth $z = z_a$.} %
The general quality ratio was defined in \cite{nasholm2009} as:%
\begin{align}
	Q\triangleq {\displaystyle \sum_{z=z_n}^{z_f}\sum_{r=0}^{\infty}\sum_{\theta=0}^{2\pi} E(\myvecx)}\left/\ {\displaystyle \sum_{z=0}^{z_n} \sum_{r=0}^{\infty}\sum_{\theta=0}^{2\pi} E(\myvecx)}\right..
\label{eq:quality_measure_generalD}
\end{align}
{The measure $Q$ illustrates the ability of the transmit beam to suppress reverberations when the first scatterings are distributed within the near-field region $z\in[0,z_n]$.} % 
A related quality measure was applied in \cite{kaupang2011} for assessment of near-field echo suppression in tissue-harmonic imaging. %

%-------------------------------------------------------------------
%\section{Discussion}
%-------------------------------------------------------------------
%* Principles, relationships and generalizations\\
%* Explanation/clarification of anomalies\\

The body-wall model chosen causes severe aberration disturbance of the transmit fields. Its characteristics, as displayed in Table \ref{tab:body_wall}, may be compared to values found at body temperature in experimental measurements on human abdomen specimens done by Hinkelman \emph{et al.}\maybespace\cite{hinkelman:ab_wall_meas}, where the amplitude correlation length is within {1.3 to 2.9}\,mm, the amplitude RMS value within {2.9 to 3.5}\,dB, the time-delay correlation length within {24 to 64}\,ns, and the time-delay correlation length within {3.3 to 17}\,mm. Especially the time-delay RMS fluctuation is more severe for the aberrative model utilized here than for the cited measurements. %

The RMS magnitude of the time-delay fluctuations introduced by the utilized aberrating layer model are, as indicated in Table \ref{tab:body_wall}, $450$\,ns, while the accumulated time-delay between $s_+(\vr,t)$ and $s_-(\vr,t)$ needed to generate maximum SURF imaging gain following the model introduced in\maybespace\cite{nasholm2009}, is smaller: $\frac{1}{2f_\HH} = 142$\,ns. The SURF synthetic transmit-beam is however shown to be generated despite these great delays induced by the body-wall. Two factors explain this: %

{1}) The $s_+$ and $s_-$ propagate following the same path and therefore experience the same aberration time-delay at each spatial point. Therefore {this abberation delay }is canceled when forming the difference field $s_\Delta(\vr,t)=s_+(\vr,t)-s_-(\vr,t)$. %

\begin{figure}[!tb]
\renewcommand{\figurename}{Fig.} % Make figure captions say Fig. instead of Figure.
	\centering
  	\includegraphics[width=\maybehalffigure]{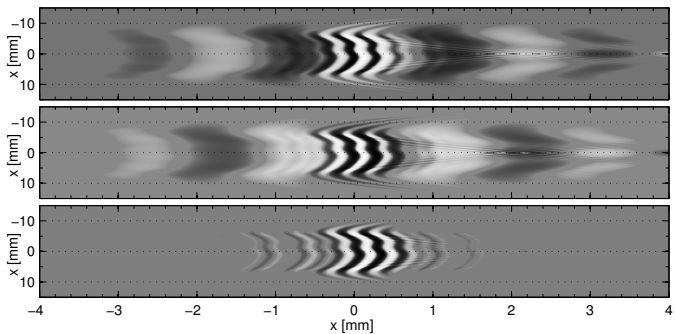}
	\caption{Propagating SURF pulse complexes sampled in the ($zx$) plane at the depth $z=6$\,mm. Horizontal axis: time, vertical axis: lateral direction. Top and middle panes: total SURF pulse complexes with opposite LF polarities. Bottom pane: extracted HF difference pulse $s_\Delta(x,t)$.}
	\label{fig:total_pulse_samples}
\end{figure}

{2}) The LF part of each SURF pulse complex experience the same aberration delay during propagation. %
Therefore the HF pulse remains within roughly the same LF pressure as when propagating without aberration. This is illustrated in Fig.~\ref{fig:total_pulse_samples}, where the total propagating SURF pulse complexes with positive and negative LF polarities are sampled at one depth within the body-wall. %

The SURF transmit-beam in the homogeneous medium has suppressed focal sidelobes compared to the fundamental transmit-beam, while the mainlobe stays unchanged. The suppression is only around $1$\,dB for the first sidelobe, while it increases to around $3$\,dB for the second sidelobe. In the inhomogeneous medium case, the sidelobes are also suppressed for SURF compared to fundamental, however about half as much as for {the} homogeneous medium. % 

Figure~\ref{fig:scores} indicates that, both for homogeneous and inhomogeneous medium, the time-shift adjustment gives superior, more robust and more predictable specific reverberation suppression quality ratio $Q_{z_a}$ than the filter-shift adjustment for depths where $z_a{>}8$\,mm, while the ratio is higher for filter-adjustment for shallower depths. The time-shift adjusted beams have $Q_{z_a}$ around $17$\,dB above $Q_{z_a}$ for the non-adjusted SURF beam, both for homogeneous and inhomogeneous medium. %

The $Q_{z_a}$ ratio calculated in the homogeneous medium for the time-shift adjusted beams, the non-adjusted SURF beams and the fundamental beams align well with what was found for the same 2-D axisymmetric simulations in \cite{nasholm2011}. The filter-adjusted SURF beams do however show worse and more unstable $Q_{z_a}$ in the present 3-D simulations than in the referred work. This is due to the averaging over the region $S$ which is used here on both $s_+(\vr,t)$ and $s_-(\vr,t)$ before $\hh_{z_a}$ is calculated. Then the canceling of temporal parts of the pulses which vary the most over the region $S$, that is the edge wave parts of the pulse, are less suppressed. The amplitude of these edge waves may be decreased if apodization is applied to the surface of pulse transmission\maybespace\cite{harris:transient_field-81, tabei:JASA-03}. That way the lateral region where $\hh_{z_a}$ generates high suppression might get larger. %

For $z_a\gtrsim8$\,mm, $Q_{z_a}$\:is lower both for the homogeneous and inhomogeneous media for time-shift, compared to for filter-adjustment. %
Therefore use of time-shift is preferred when $z_a$ is {far from} the transducer. %
{This is also reflected in the $z_a=39$\:cm section of Fig.~}\ref{fig:adjusted_beams} both with and without the body-wall. There significant in-focus field degradation is observed for the filter-adjustment beam as opposed to for the time-shift adjusted beam. %

{%
Ultrasound images of patients whose anatomy cause noisy images, for example due to obesity, are likely to be blurred owing to both aberration and reverberation noise. }%

{The repetitive nature of reverberation noise makes the receive signals correlated both in the temporal and the spatial directions. This may obstruct estimation of the filter that is used in aberration correction schemes like }\cite{masoy:JASA-05}. %
By suppression of reverberation noise, estimation of the aberration correction filter is thus enhanced, and therefore the aberration correction itself is likely to get more accurate. %
Instruments that may combine aberration correction with reverberation suppression are thus, compared to instruments with aberration correction alone, likely to produce images of enhanced quality also regarding the wave-front aberration noise. %

%
%- Speculation about research that might be done in the future to build upon results in the present paper. NO remarks about what authors intend to do next:

    %* Definition of conditions under which conclusions apply\\
    %* Agreement or disagreement with previous work\\
    %* Theoretical implications\\
    %* Possible applications\\
    %* Further thought or research\\
    %* Summary/conclusion -- significance of work\\

%-------------------------------------------------------------------
\section{Concluding remarks}
%-------------------------------------------------------------------
This work indicates through computer simulations that generation a SURF reverberation suppression synthetic transmit-beam is feasible in spite of the propagating waves being distorted by an inhomogeneous medium which emulates a strongly aberrating body-wall. Generation of the beam is hence also likely to be possible when utilizing body-wall models producing less severe aberrations corresponding to what was measured {e.g.} in \cite{hinkelman:ab_wall_meas}. %

The post-processing suppression depth adjustment methods, as previously implemented in \cite{nasholm2011} for a homogeneous medium, are here shown to be feasible on the simulated aberrated fields {also} within an inhomogeneous medium. %for the purpose of being able to shadow noise due the first multiple scattering by some especially strong object or interface located at the depth $z_a$. %

%The benefit of utilizing the general operator-shift processing method instead of the time-shift based one is largest for shallow depths also . %
%
%The conclusions regarding the reverberation suppression abilities are based on synthetic SURF transmit-beams after adjustment. 

%A strength of the method is that%
%

%(- Not repeat abstract, or subject\\
%- Review principal results, and where they emerged\\
%- Broad discussion about possible implications\\

Further research of interest within the field also includes \emph{in vivo} imaging through true body-walls using the investigated methods and also performance comparisons with tissue harmonic imaging, especially when utilizing pulse inversion. %A first step is to perform the tests in a water-tank using some aberrating medium within the first part of the propagation path, as well as imaging tests on aberration-emulating phantoms. %

{Non-adjusted and adjusted beams could also be generated from SURF transmit fields measured by a hydrophone in a water-tank. Beam generation with a inhomogeneous body-wall setup may be emulated using some aberrating material within the tank, \emph{e.g}\ as done by use of gel layers in }\cite{jing:JASA-07}. %
It is furthermore desirable to perform{ }numerical propagation simulations using a tool that properly handles multiple scattering (as does not {the} forward-propagation method used here) in order to further quantify the reverberation suppression ability of the methods within an aberrative medium, as well as doing comparisons to the performance of tissue harmonic imaging using the same setup. %

{We expect the family of SURF reverberation suppression methods to come out as versatile techniques to enhance ultrasound image reconstruction also in clinical settings.}
%For an ultrasound imaging system capable of transmitting dual-frequency SURF pulses, both 
%
%

\bibliographystyle{IEEEtran}
%{\footnotesize\bibliography{SURF}}
%{\bibliography{SURF}}


\begin{IEEEbiography}[%
	{\includegraphics[width=1in,height=1.25in,clip,keepaspectratio]{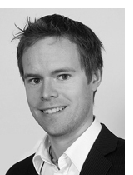}}]{Sven Peter Näsholm}\footnotesize 
was born in Örnsköldsvik, Sweden, in 1975. He received his M.Sc. degree in engineering physics in 2002 from Umeå University, Sweden. %
In 2008, he successfully defended his Ph.D. thesis entitled ``Ultrasound beams for enhanced image quality'' at the Norwegian University of Science and Technology, Trondheim, Norway. %
In 2009, he joined the Digital Signal Processing and Image Analysis group as a pos-doctoral fellow at the Department of Informatics, University of Oslo, Norway. %
His research interests are within the fields of sonar and ultrasound imaging including nonlinear effects, transducer design, field simulation, and acoustic noise suppression.
\end{IEEEbiography}
\begin{IEEEbiography}[%
{\includegraphics[width=1in,height=1.25in,clip,keepaspectratio]{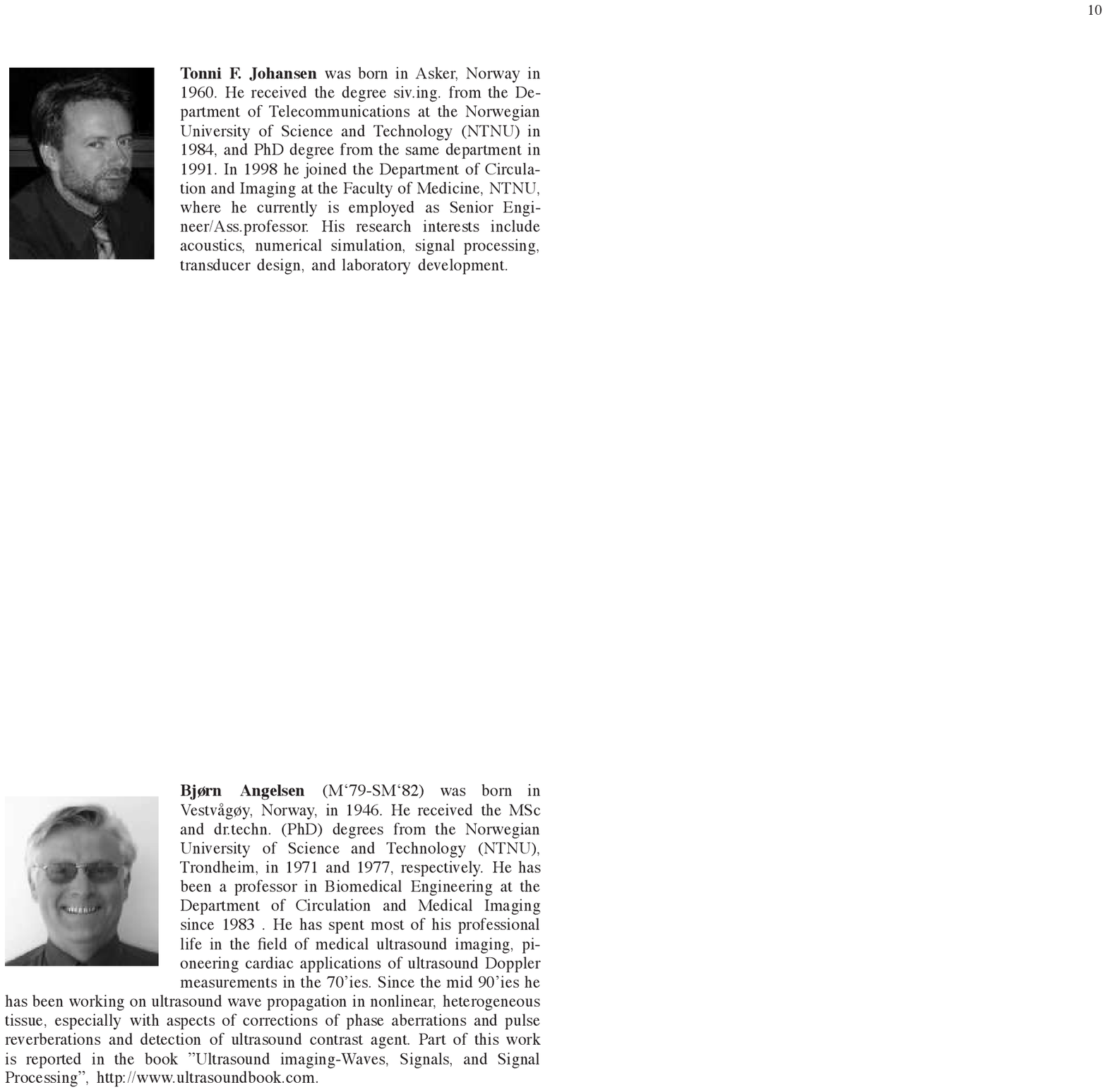}}]{Bjørn A. J. Angelsen}\footnotesize
was born 1946 in Vestvågøy, Norway. He received a MEE in 1971 from the Norwegian University of Science and Technology, Trondheim, and a Ph.D. from the same University in 1977. His Ph.D. work was on Doppler ultrasound measurement of blood velocities and
flow in the heart and the large arteries. He is a professor of Medical Imaging at the same university since 1983. In 1977--78 he was a visiting Post Doc at University of California, Berkeley, and Stanford Research Institute, Palo Alto. He has been strongly involved in development of cardiac ultrasound imaging instruments in collaboration with Vingmed Ultrasound, now GE Vingmed Ultrasound. He has written textbooks on ultrasound cardiac Doppler measurements and theoretical ultrasound acoustics, and holds several patents in the field of ultrasound imaging.
\end{IEEEbiography}
\vfill
\end{document}